\documentclass[
 reprint,
superscriptaddress,
 amsmath,amssymb,
 aps,
]{revtex4-2}
\usepackage[dvipsnames]{xcolor}
\usepackage{graphicx}
\usepackage{dcolumn}
\usepackage{graphicx}
\usepackage{media9}
\usepackage{physics}
\usepackage{amssymb}
\usepackage{amsmath}
\usepackage{xcolor}
\usepackage[T1]{fontenc}
\usepackage[latin9]{inputenc}
\usepackage{soul}
\usepackage{amsmath}
\usepackage{dsfont}
\usepackage{amstext}
\usepackage[normalem]{ulem}

\usepackage{tocvsec2}

\usepackage{amssymb}
\usepackage{amsbsy}
\usepackage{amsthm}
\usepackage{epsfig}
\usepackage{framed}
\usepackage{graphicx}
\usepackage{bbm}
\usepackage{hyperref}
\usepackage{color}
\usepackage{multirow}
\usepackage{changes}
\usepackage{bm}

\makeatletter

\newcommand{\Rmnum}[1]{\expandafter\@slowromancap\romannumeral #1@}
\makeatother

\newcommand{\bea}{\begin{eqnarray}}
\newcommand{\eea}{\end{eqnarray}}
\newcommand{\bpm}{\begin{pmatrix}}
\newcommand{\epm}{\end{pmatrix}}
\newcommand{\bal}{\begin{aligned}}
\newcommand{\eal}{\end{aligned}}

\usepackage{babel}
\usepackage{changes}
\begin{document}

\preprint{APS/123-QED}

\title{Ultrafast laser-driven dynamics in metal-insulator interface
}

\author{Abdallah AlShafey}
\email{alshafey.1@osu.edu}
\affiliation{Department of Physics, The Ohio State University, Columbus, Ohio 43210, USA}
\author{Gerard McCaul}
\affiliation{Tulane University, New Orleans, Louisiana 70118, USA}
\author{Yuan-Ming Lu}
\affiliation{Department of Physics, The Ohio State University, Columbus, Ohio 43210, USA}
\author{Xu-Yan Jia}
\affiliation{Department of Physics, Beihang University, Beijing 100191, China}
\author{Shou-Shu Gong}
\affiliation{Department of Physics, Beihang University, Beijing 100191, China}
\author{Zachariah Addison}
\affiliation{Department of Physics, The Ohio State University, Columbus, Ohio 43210, USA}
\author{Denys I. Bondar}
\affiliation{Tulane University, New Orleans, Louisiana 70118, USA}
\author{Mohit Randeria}
\affiliation{Department of Physics, The Ohio State University, Columbus, Ohio 43210, USA}
\author{Alexandra S. Landsman}
\email{landsman.7@osu.edu}
\affiliation{Department of Physics, The Ohio State University, Columbus, Ohio 43210, USA}
\date{\today}
\begin{abstract}

The nearly free electron metal next to a localized Mott insulating state has been recently proposed as a way to probe Kondo lattice physics and to gain insight into how the two extremes of localized and delocalized electron states interact (Sunko, et al, Science advances 6, 2020). Although high harmonic generation has been used extensively to investigate the gas phase, its extension to solids is relatively recent, and has not yet been applied to interfaces. Here, we investigate the field-induced dielectric break-down at the Mott-insulator/metal interface using high harmonic generation, emitted when the interface is subjected to an ultrafast laser pulse. We show that the intensity of high harmonic emission correlates closely with doublon production and the corresponding loss of short-range anti-ferromagnetic order. For strong interlayer coupling, the harmonic intensity is defined by a phase transition between states that do not exist in a pure Mott insulator case. For weak interlayer coupling, the threshold for dielectric breakdown is considerably lowered due to the presence of a metallic layer. This suggests that interlayer coupling can be used as an additional knob to control magnetic insulator break-down, with implications for using Mott insulators as memristors in neuromorphic circuits.

\end{abstract}

\maketitle

The creation and efficient control of material interfaces has always been pivotal in the development of new technology. From the basic transistor to the ongoing pursuit of ultrafast optical control of magnetic states in spin electronic devices \cite{Siegrist:19, hellman2017interface, kirilyuk2010ultrafast}, the interaction between materials of different electronic properties at the interface consistently introduces novel and useful physics. Of particular importance is the interface between a correlated insulator and metal, which provides a setting for the interaction between the localized and delocalized extremes of electron motion in a solid. Metal-insulator (MI) interfaces have seen applications in several systems such as in the development of memristor technology \cite{strukov2008missing}, control of magnetic properties through spin pumping across the interface \cite{heinrich2011spin}, and others.
As such, it is of great importance to develop both a theoretical understanding and an experimental probe for the charge dynamics in such systems.

Parallel to these developments has been the discovery and application of High Harmonic Generation (HHG). This phenomenon was first reported in atomic gases \cite{ferray1988multiple}, where the optical response of the gas produced frequencies many multiples greater than that of the input light, and was subsequently utilized in the development of attosecond science \cite{Ciappina2015PRL,li2020attosecond}. More recently, HHG has been observed in semiconductors \cite{ghimire2011observation, hohenleutner2015real, luu2015extreme} and nanostructures \cite{Lisa2017PRL,Vampa2017}, accompanied by  theoretical development of the mechanism for HHG in solids \cite{vampa2014theoretical, ghimire2019high}. Currently, there is mounting interest in the production of HHG and the accompanying ultrafast charge dynamics in strongly correlated materials \cite{Silva:18, murakami2021high, orthodoxou2021high}, for which the single-particle picture breaks down. The applications of this effect are numerous, ranging from quantum control \cite{tracking1,tracking2} to single atom computing \cite{2104.06322}.
Of particular relevance to this present work is the use of HHG as a spectroscopic probe for not only different phases in a solid \cite{silva2019topological}, but also material characterisation and identification \cite{mixing}. 

In this Letter, we combine these avenues of investigation, and study an optically driven metal-magnetic insulator interface. We find that the presence of the interface has a significant impact on both optical response and internal spin dynamics of the system, where a critical strength of interlayer coupling demarcates different dynamical phases with distinct response profiles. Additionally, we find that peak optical emission of this system is well correlated with doublon density, indicating that optical response can be used to infer critical information on the state of the system.  The novel behaviour of such driven interfaces suggest a number of potential applications, including high precision metrology and logic components.

\begin{figure}[b]
\includegraphics[width=.48\textwidth]{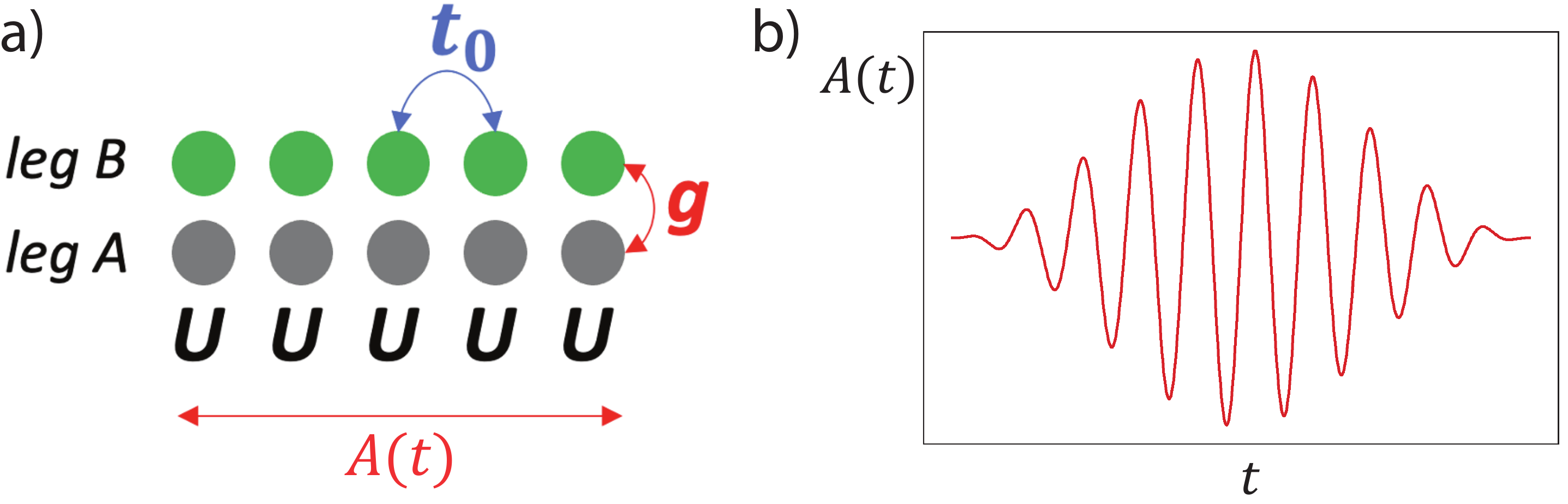}
\caption{\label{fig:system} \textbf{a)} Model system with hopping parameters $t_0$ and $g$, and Hubbard repulsion $U$. Leg $A$ models the correlated insulator layer and leg $B$ models the metallic layer.  \textbf{b)} Vector potential $A(t)$ with duration $N=10$ periods, carrier frequency $\omega_0/t_0=0.262$, and peak field strength $F_0=0.769t_0/a$, where $a$ is the lattice constant.}
\end{figure}

We wish to understand the characteristics of both the strong field optical response and internal dynamics of a metal-Mott insulator interface, and how this might differ from a pure Mott insulator investigated previously in a single Hubbard leg with half-filling \cite{PhysRevB.86.075148,Silva:18,mccaul2020driven}.  For the pure Mott insulator case, dielectric breakdown, accompanied by high harmonic emission and the loss of short-range antiferromagnetic ordering, is observed when the driving field strength exceeds a critical value given by \cite{PhysRevB.86.075148,Silva:18,mccaul2020driven}, 
\begin{equation}
    F_{\rm{crit}}\sim\frac{\Delta}{2\xi}, \label{eq:threshold}
\end{equation}
where $\Delta$ is the Mott gap \citep{PhysRevB.79.245120}, and $\xi$ is the doublon-hole correlation length \citep{PhysRevB.86.075148}. 

We find that the presence of weak coupling across layers facilitates doublon-holon pair production, leading to a lower threshold value of $F_{\rm crit}$, and explain this as resulting from increased correlation length, $\xi$.  On the other hand, in the presence of strong interfacial coupling, we find that increasing $g$ leads to an increase in the effective bandgap, $\Delta$ (see Eq.~\ref{eq:threshold}), leading to an increase in the threshold value, $F_{\rm crit}$, needed to create doublon-holon pairs.  

We employ the asymmetric two-leg ladder depicted in Fig.~\ref{fig:system} as a model for a metal-insulator interface. The static model has been used in \cite{Takashi:20} to study the Kondo lattice physics of $\text{PdCrO}_2$, a magnetic oxide metal with a layered structure of alternating metallic and Mott insulating layers.
This model consists of an insulating layer labeled as $A$ with onsite interactions,  which is coupled to the free leg $B$ which models the metallic layer across the interface.  The kinetic term on each leg is parametrised by the intrachain hopping parameter $t_0$, while the coupling strength between the chains is given by the hopping $g$. The  onsite repulsion on the $A$ leg is parametrised by $U$, and provides the simplest introduction of repulsive interactions between the electrons.  The single-leg Hubbard chain has been used previously to investigate HHG in strongly correlated systems \cite{Silva:18, murakami2021high, orthodoxou2021high, mccaul2020driven}. 

Prior theoretical study of the undriven system shown in the left panel of Fig.~\ref{fig:system} found rich behavior with different states that depend on the values of on-site repulsion and the two hopping parameters.  In particular, in addition to the Luttinger liquid phase present in a single Mott insulator chain, prior work \cite{abdelwahab2015ground, abdelwahab2018correlations} found {\it Kondo-Mott insulator, spin-gapped Mott insulator and correlated band insulator states}, with successive onset corresponding to increasing values of the inter-layer coupling parameter, $g$, respectively.  Focusing in particular on low and high values of $g$, we find a qualitatively different relationship between onsite interaction, $U$, and the strength of high harmonic emission, suggesting that HHG can be used as a phase-sensitive spectroscopic probe of correlated electron dynamics at interfaces.  Moreover, we demonstrate a clear correlation between the intensity of the nonlinear optical response, doublon production and the loss of short-range anti-ferromagnetic order that accompanies the break-down of the Mott insulator.

In our model, we introduce a dipolar coupling to an electric field driving both metal and insulator parallel to the leg direction. The Hamiltonian describing this system is:
\begin{eqnarray}
\label{eq:model}
\hat{H}(t)=-t_0&&\sum_{\alpha,\sigma}\sum_{r=1}^{L_x}\; {\rm e}^{-i\phi(t)}c_{r,\alpha,\sigma}^{\dagger}c_{r+1,\alpha,\sigma}+\textrm{h.c.}\nonumber\\-g&&\sum_\sigma\sum_{r=1}^{L_x}\;c_{r,A,\sigma}^{\dagger}c_{r,B,\sigma}+\textrm{h.c.}\nonumber\\
+U&&\sum_{r=1}^{L_x}\;\Big(n_{r, A, \uparrow}-\frac{1}{2}\Big)\Big( n_{r, A, \downarrow}-\frac{1}{2}\Big),
\end{eqnarray}
where $c_{r,\alpha}$ ($c_{r,\alpha}
^\dagger$) is the annihilation (creation) operator on site $r$, respectively, $n_{r,\alpha}=c_{r,\alpha}^\dagger c_{r,\alpha}$ is the local density operator, $\alpha = \{A,B\}$ sums over the Hubbard leg $A$ and free leg $B$ respectively, while $\sigma = \{\uparrow, \downarrow\}$ sums over the electron spins. The  dipolar coupling to a short strong-field pulse is incorporated via  the Peierls substitution $t_0\rightarrow t_0 {\rm e}^{-i\phi(t)}$, with $\phi(t)=aA(t)$. Here $A(t)$ is the driving field vector potential given by a transform limited field described by:
\begin{equation}
A(t)=\frac{F_0}{\omega_0}\sin^2{(\frac{\omega_0 t}{2N})}\sin{(\omega_0 t)}. 
\end{equation}
Unless otherwise stated, the amplitude and frequency parameters are set as $F_0=10$~MV/cm with frequency $\omega_0=32.9$~THz \citep{Hohenleutner2015,Silva:18}. The choice of laser parameters puts us comfortably in the strong field tunnelling regime corresponding to the Keldysh parameter $\gamma\equiv \hbar\omega_0/\xi F_0\ll 1$, where $\xi$ is the doublon-hole correlation length \cite{floriangebhard2010}. Finally, simulations are performed via exact diagonalization in QuSpin \cite{quspin}, evolving from the ground state at half filling (i.e. $\sum_{r,\alpha,\sigma} n_{r,\alpha,\sigma}=2L_x$) and periodic boundaries in the direction, with $L_x=6$.

The optical response of this system (via the Larmor formula) will be proportional to $\frac{\partial J(t)}{\partial t}$, where $J(t)=\langle{\hat{J}(t)}\rangle$ is the current expectation. The current operator is itself defined via a continuity equation, and the contribution from each leg is
\begin{equation}
    \hat{J}_\alpha(t) = -iat_0\sum_{\sigma}\sum_{r=1}^{L_x}{\rm e}^{-i\phi(t)}c^{\dagger}_{r,\alpha,\sigma}c_{r+1,\alpha,\sigma}-\textrm{h.c.},
\label{eq:current}
\end{equation}
such that $\hat{J}=\hat{J}_A + \hat{J}_B$.
The spectral response of the current, corresponding to the intensity of high harmonic generation (HHG), is given by: $S(\omega)=\mathcal{F}\left[\frac{\partial J(t)}{\partial t}\right]$ . 

\begin{figure*}
\centering
\includegraphics[width=.99\textwidth]{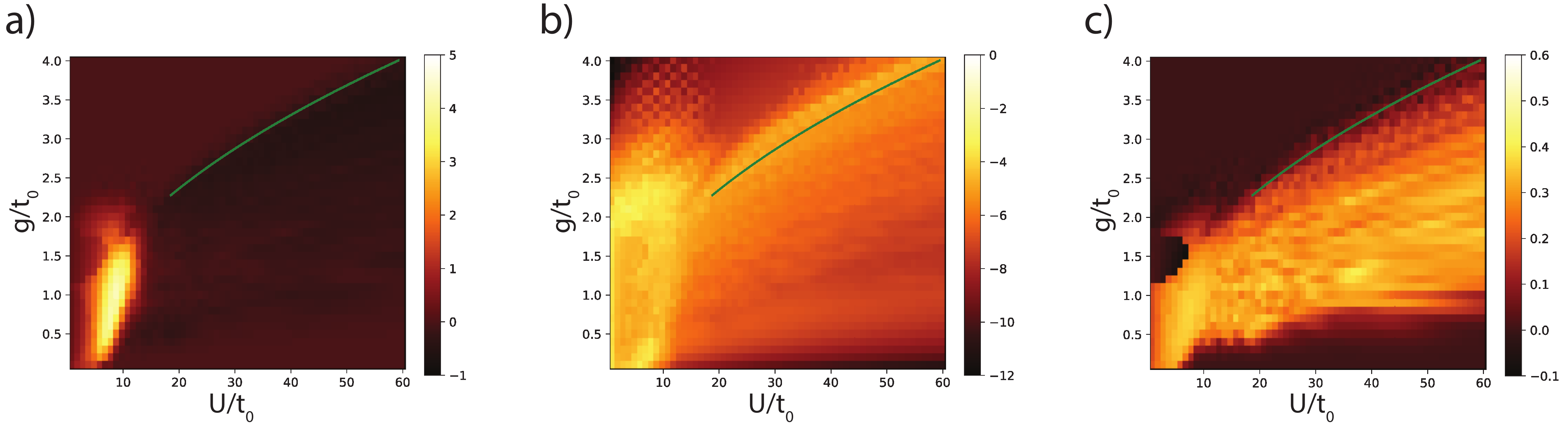}
\caption{\label{fig:maxes}  Color plots showing the response of the Hubbard leg: \textbf{a)} the change in doublon density $\delta D$ (Eq.~\ref{eq:DDD}), \textbf{b)} the maximum of the spectral response $\max_\omega\left[S(\omega)\right]$ on a log-scale, and \textbf{c)} the change in nearest-neighbor spin-spin correlations $\delta \eta$, all as functions of $U/t_0$ and $g/t_0$.  The green line separates the spin-gapped Mott insulator from the correlated band insulator phases.}
\end{figure*}

Importantly, the emission of high harmonics is accompanied by the break-down of short-range anti-ferromagnetic ordering in the Mott chain, resulting in doublon-hole pair production, as shown in Fig.~\ref{fig:maxes}.  
The magnitude of this breakdown is reflected in the average doublon density given by \citep{PhysRevB.86.075148}
\begin{equation}
    D_\alpha(t) = \frac{1}{L_x}\sum_{r=1}^{L_x}\;\langle n_{r, \alpha, \uparrow}(t) n_{r, \alpha, \downarrow}(t)\rangle.
\label{eq:D}
\end{equation}
Hence when the critical field amplitude of the driving laser field is exceeded, we expect to see an \textit{increase} in this density over the course of the evolution. Defining $\delta D$ as a measure of this change,
\begin{equation}
    \delta D = \frac{1}{2}\sum_\alpha \frac{D_\alpha(t_f)-D_\alpha(0)}{D_\alpha(0)},
\label{eq:DDD}
\end{equation}
we expect $\delta D >0$ when $F_0>F_{\rm crit}$, where $F_{\rm crit}$ is the minimum amplitude of the laser field required to observe high harmonics \cite{Silva:18}.

A closely related quantity is the nearest neighbor spin-spin correlation \begin{equation}
    \eta_A(t) = \frac{1}{L_x}\sum_{r=1}^{L_x}\;\langle \textbf{S}_{r, A}(t)\cdot \textbf{S}_{r+1, A}(t)\rangle,
\label{eq:eta}
\end{equation}
which initially would be negative because of the anti-ferromagnetic nature of the unperturbed Hubbard chain, and would increase with time as doublon-hole pairs are produced and short-range anti-ferromagnetic ordering breaks down. We can define $\delta\eta$ similarly to Eq.~\ref{eq:DDD}.
As Fig.~\ref{fig:maxes} shows, there is a close correlation between all three of the above defined quantities, $S(\omega)$, $\delta D$ and $\delta\eta$.

\begin{figure*}
\includegraphics[width=.8\textwidth]{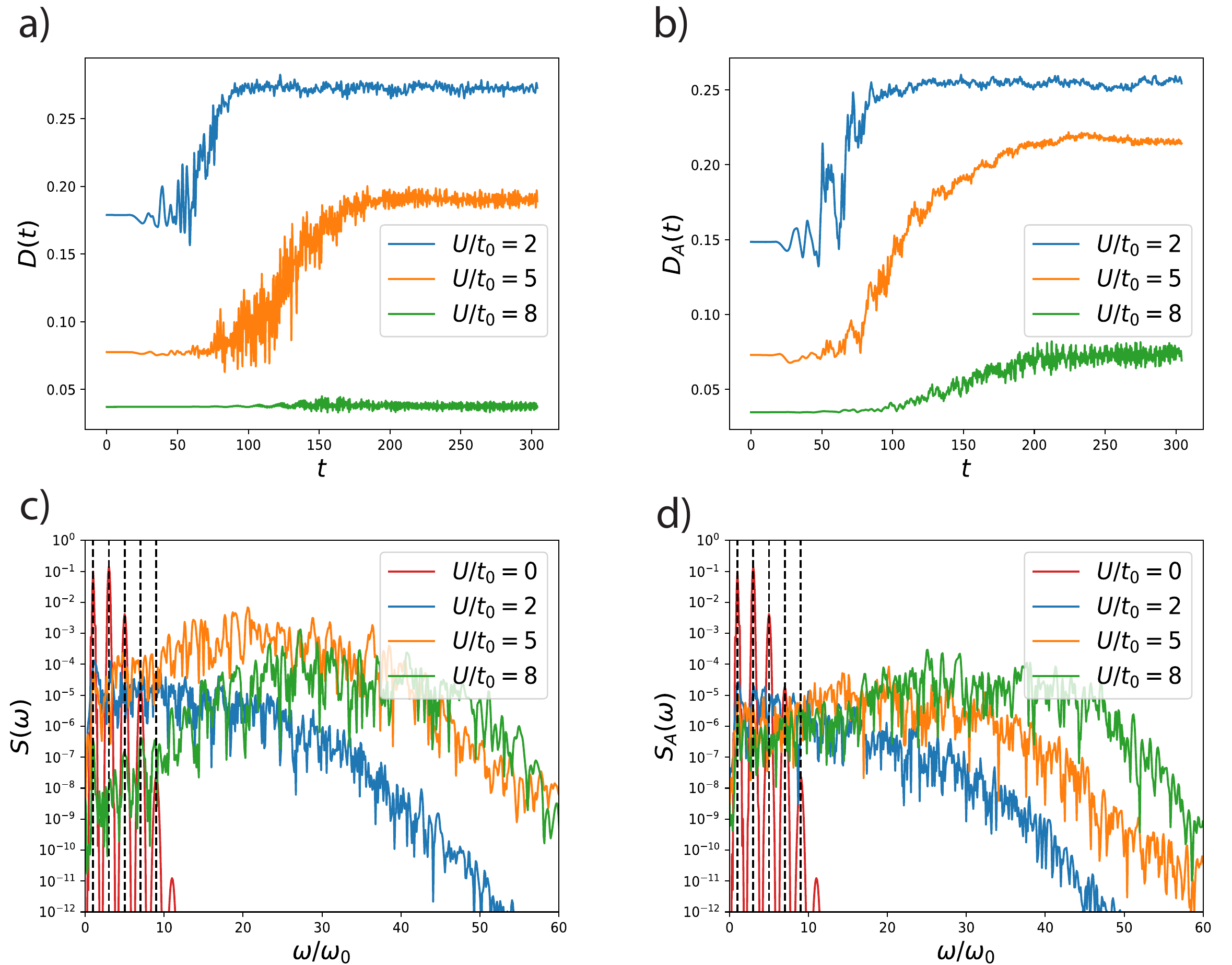}
\caption{\label{fig:spectra}  \textbf{a, c)} $D(t)$ and $S(\omega)$ for uncoupled chain $g/t_0=0$. \textbf{b, d)} $D_A(t)$ and $S_A(\omega)$ showing the response of the Hubbard leg for finite coupling $g/t_0=0.2$. Note how $U/t_0=8$ shows breakdown for $g/t_0=0.2$ in contrast to the uncoupled case, as indicated by an increase in $D(t)$. This response can be experimentally measured as a significant increase in the low-order HHG spectrum (bottom panels). The red HHG plot is the response for $U/t_0=0$ case, used for reference.}
\end{figure*}

\begin{figure}
\centering
\includegraphics[width=.45\textwidth]{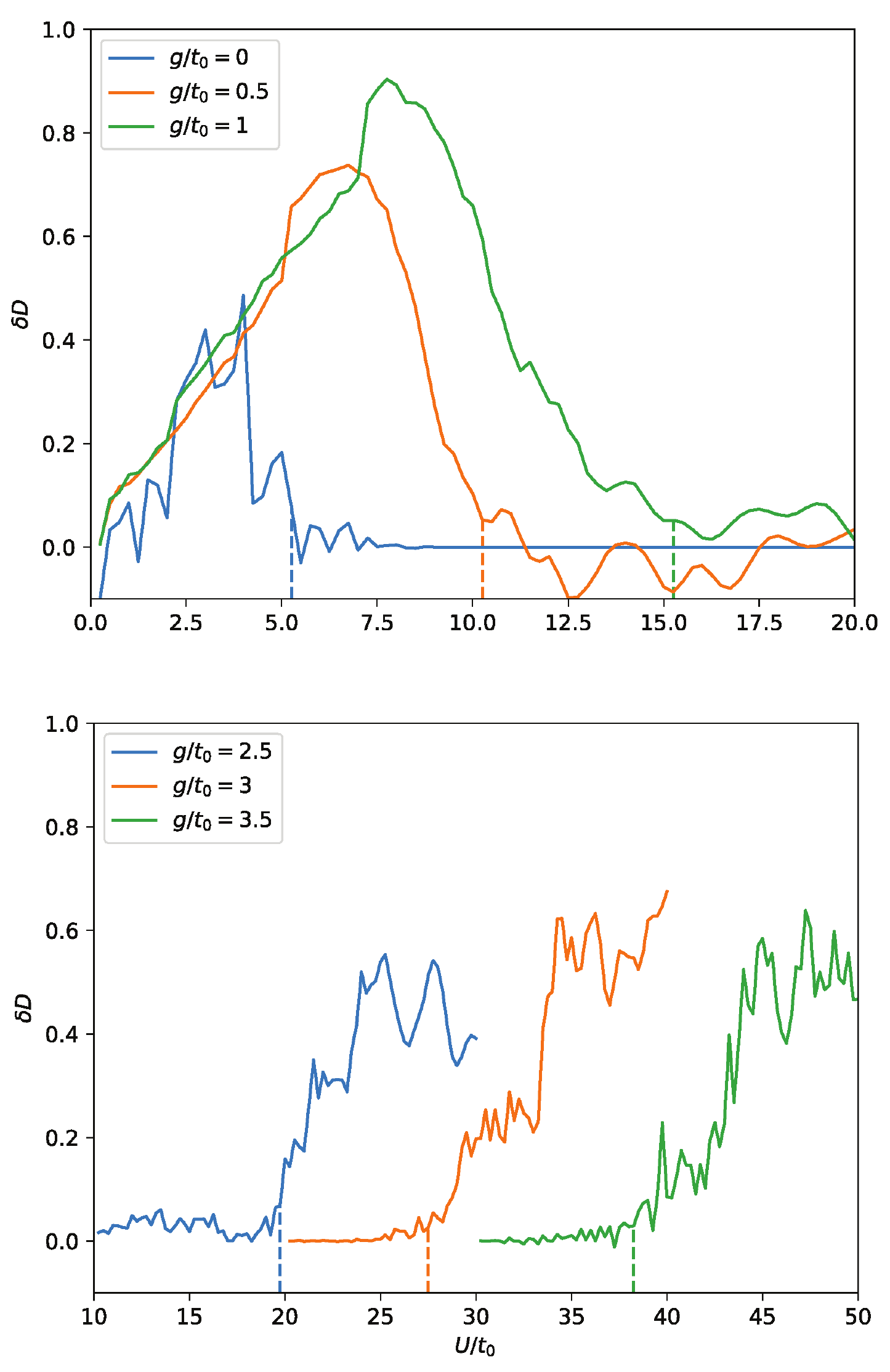}
\caption{\label{fig:deltaD} Change in doublon density over the course of the system evolution. For small $g$, doublon production occurs at much larger values of $U$ compared to the uncoupled case. When $g>g_{\rm crit}$, the doublon behaviour changes dramatically, and there is a \textit{minimum} $U$ required for doublon production. }
\end{figure}

For an uncoupled Hubbard chain, corresponding to $g=0$, a broad white light-like spectrum is produced where the peak of the spectrum shifts to higher harmonics, $N$, according to $N\sim \frac{U}{\omega_0}$ \cite{Silva:18}. This uncoupled case is shown in the left panels of Fig.~\ref{fig:spectra}. Here we see the expected combination of the free  leg's well defined peaks at odd harmonics together with the broader spectrum of the Hubbard leg. Introducing a small interaction $g=0.2t_0$ for Fig.~\ref{fig:spectra} (right panels) significantly alters the overall spectral response. This is particularly apparent at lower harmonic orders, where harmonic intensity becomes comparable for all plotted values of $U/t_0$.  Note that the harmonic intensity is plotted on a log scale, showing orders of magnitude increase in harmonic intensity for the highest $U$ case (corresponding to $U/t_0=8$).  The increase in harmonic intensity due to the interface coupling term is also reflected in the doublon production (compare green curves in the left and right panels of Fig.~\ref{fig:spectra}).  Overall, it is clear that the presence of the interface lowers the threshold for doublon production and the corresponding break-down of the Mott insulator.  This is also clear from the top panel of Fig.~\ref{fig:deltaD}, where increasing values of interfacial coupling $g$ leads to an overall increase in doublon production, as well as a shift to larger values of $U$ where this production and the corresponding HHG intensity is maximized.

\begin{figure}[b]
\includegraphics[width=.5\textwidth]{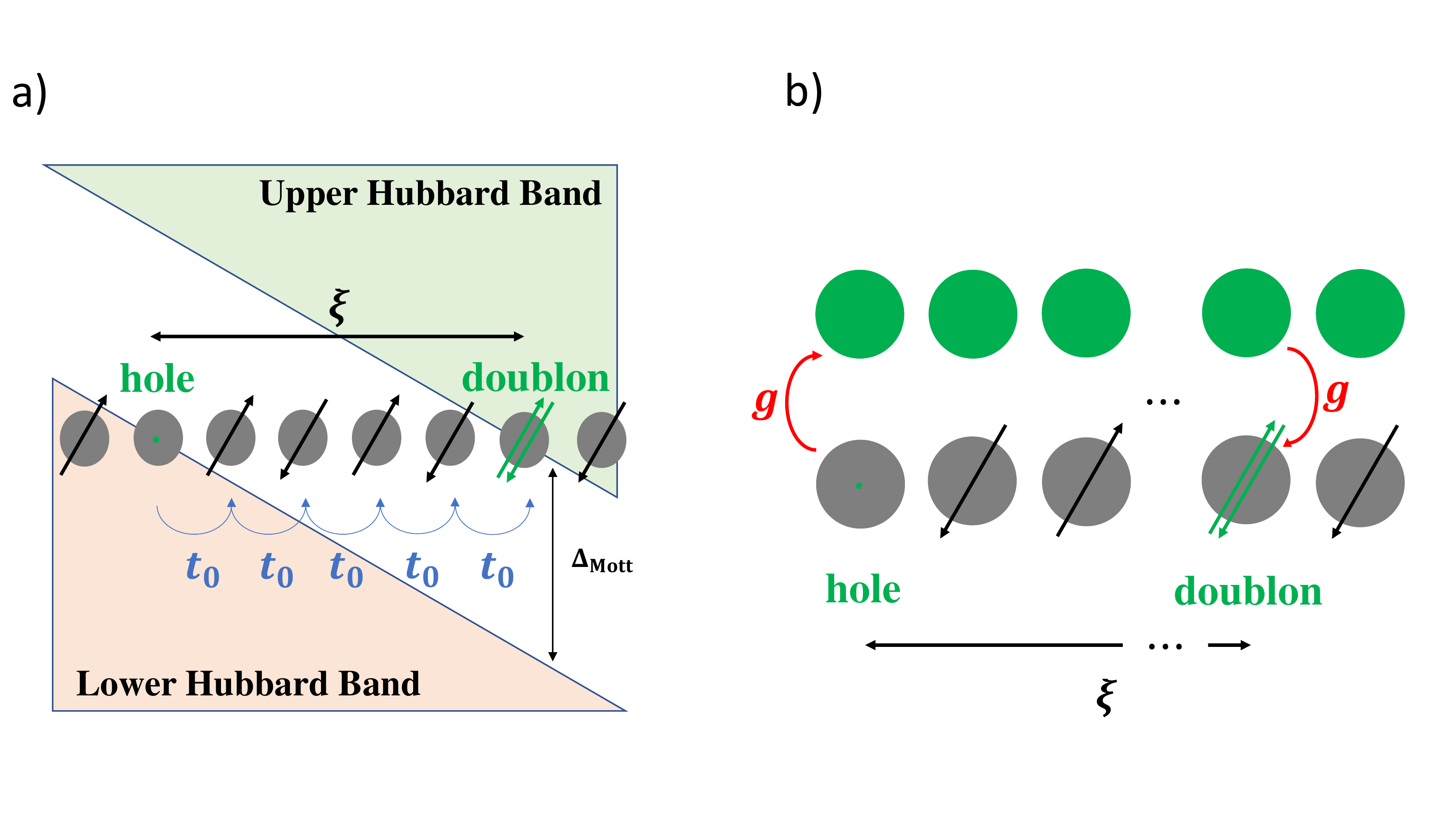}
\caption{\label{fig:newmech} \textbf{a)} In the uncoupled chain, we rely on many $t_0$ hoppings to create a pair of size $\xi$. \textbf{b)} In the coupled system, a doublon-hole pair of arbitrary size can be created by just two hoppings $g$.}
\end{figure}

This dramatic enhancement in optical response due to the presence of a metallic layer can be interrogated by considering how the correlation length $\xi$ is modified by the introduction of the interlayer coupling $g$.  In particular, longer $\xi$ will lower the critical driving amplitude for dielectric breakdown, per Eq.~\ref{eq:threshold}.  
To calculate $\xi$, we performed DMRG (density matrix renormalization group) calculations
on the $2 \times 64$ ladder with the open boundary conditions. We kept the 4096 SU(2) multiplets, discarding weights smaller than $10^{(-7)}$.  These calculations show that the correlation length increases substantially with interfacial coupling, $g$ (see SI for table of results).  This can be understood via Fig.~\ref{fig:newmech}, as the interface provides a new mechanism for the creation of doublon-hole pairs. In the single chain a pair with separation $ka$ would requite $k$ hoppings with amplitude $t_0$, with a rate of doublon production $\Gamma\sim(t_0/U)^k$ in the $U\gg t_0$ limit \cite{oka2012nonlinear}. 
On the other hand, the creation of a doublon-hole pair in the coupled metal-insulator system can in principle be achieved with just two $g$ hoppings. This explains why the presence of the interface dramatically increases the correlation length, as observed in our DMRG simulations.

The above analysis is fully compatible with the results shown in the upper panel of Fig.~\ref{fig:deltaD}, where increasing $g$ from zero leads to a dramatic increase in doublon production,
going to substantially higher values of $U$.  This is accompanied by an increase in correlation length, $\xi$, and hence a reduction in the critical field amplitude, 
$F_{\rm{crit}}$, see Eq.~\ref{eq:threshold}. Note the contrast to the uncoupled insulator, where the chosen field amplitude $F_0$ is only sufficient to cause dielectric breakdown up to $U\approx 6.8t_0$. 

A dramatically different response occurs at high values of $g$, corresponding to the region around the correlated band insulator and the spin-gapped Mott insulator phases \cite{abdelwahab2015ground, abdelwahab2018correlations}.  In particular, the transition between these two phases occurs at a critical value of $g$, given by $g_{\rm crit}=2t_0$ in the limiting case of $U=0$ and going to higher $g_{\rm crit}$ values with increasing $U$ (see Figure 1 of \cite{abdelwahab2018correlations}).  A striking response, which corresponds to a transition from the correlated band insulator to the spin-gapped Mott insulator phase with increasing value of $U$, is shown in the lower panel of Fig.~\ref{fig:deltaD}.  In particular, for a given value of $g$, there is a minimum value of $U$ above which doublon production takes place.  For instance, for $g/t_0 = 2.5$, this minimum value is given by $U_{\rm min}/t_0 \approx 20$, as can be seen in the lower panel of Fig.~\ref{fig:deltaD}.

This is of course the opposite of the small $g$ case, shown in the upper panel of Fig.~\ref{fig:deltaD}, where for any $g$ value, there is a response for small $U$, which disappears as $U$ increases above a certain threshold, $U_{\rm crit}$, where the value of $U_{\rm crit}$ increases with $g$.  Hence for values of $g/t_0$ corresponding to $0$, $0.5$ and $1$, $U_{\rm crit}/t_0$ corresponds to roughly $5$, $10$ and $15$, respectively.

To understand the loss of spectral response to the driving laser field above $g_{\rm crit}$, let
us turn to the easily solvable $U=0$ metallic limit. In this limiting scenario it is possible to understand the sudden change in response analytically. In this case the spin index is superfluous and can be suppressed, while the Hamiltonian can be diagonalised.
One would then obtain a two band Bloch Hamiltonian with band energies given by 
\begin{equation}
    \epsilon_{\omega_k} = -2t_0 \cos(\omega_k-\phi(t)) \pm g.
\label{eq:disp}
\end{equation}
From this perspective it is clear that $g_{\rm crit}/t_0 = 2$ corresponds to the point at which a gap opens between the two bands in the non-interacting electron limit.
It is also worth noting that gapping the system kills off the response completely in this $U=0$ band limit, regardless of the strength of the light pulse. This is because the $k$-dependence in the Bloch Hamiltonian appears solely in the identity matrix term, meaning that the coupling of the light through $k\rightarrow k-eA$ becomes trivial. In particular, the dynamical equation for the density matrix $i\Dot{\rho} =[H,\rho]$ is unaffected by the introduction of the pulse.

As $U$ increases from zero, the value of $g_{\rm crit}$ marking the transition from the spin-gapped Mott insulator to the  correlated band insulator increases as well, consistent with the lower panel of Fig.~\ref{fig:deltaD}.

In the dimer limit, corresponding to $g_{\rm crit} \gg t_0$, the gap is described by (see SI for details),
\begin{equation}
\label{eq:approxgap}
\Delta \approx 2\Big(\sqrt{(2g)^2+(U/4)^2}-\sqrt{g^2+(U/4)^2}\Big)-4t_0.
\end{equation}
For large $U$, this gap can be approximated by  $\Delta\approx 12g^2/U-4t_0$, explaining the $g\propto \sqrt{U}$ dependence plotted as a green line in all panels of Fig.~\ref{fig:maxes}, and predicting a paradoxical effect that it is possible to lower the gap by \textit{increasing} $U$. Note that this green line marks a transition between the spin-gapped Mott insulator and correlated band insulator phases, demonstrating that high harmonic emission can be used as an experimental probe of phase transitions in correlated materials.

In conclusion, we show that 
the presence of a metallic layer coupled to a magnetic insulator interface, modelled by the Hubbard chain, considerably alters the nonlinear correlated electron response driven by the ultrafast laser field.  The change in underlying dynamics as the interfacial coupling $g$ increases is explained by increased doublon-holon correlation length in the weak coupling regime, and increased Mott gap in the strong coupling regime.  In the former case, the threshold electric field required to achieve dielectric breakdown is lowered, while in the latter case, strong coupling prevents doublon-holon pair production by pushing the system into the correlated band insulator phase.  We furthermore show that the $U$ and $g$-dependent phase transition from the correlated band insulator to the spin-gapped Mott insulator is accompanied by a broad plateau spectrum of high harmonics, which easily distinguishes the magnetic insulator layer response from the well-defined lower cut-off harmonic peaks coming from the metallic layer.  Across all parameters, the doublon density correlates with nonlinear optical response, suggesting that harmonic emission can be used as a direct probe of internal spin dynamics in interfacial systems, which themselves display a much richer range of driven behaviours than a pure Mott insulator.  

The existence of a critical interlayer coupling $g$ which completely eliminates the optical response, may seem surprising given that it appears to be insensitive to both the amplitude and frequency of the driving.
This binary behaviour has natural appeal as a physical substrate for logic gates, and it is possible to envisage an `optical transistor' controlled via interlayer distance, which will determine $g$. For weak interlayer coupling, the dependence of the Mott insulator breakdown on the tunable value of $g$ suggests applications to neuromorphic circuits, where the Mott insulator has been proposed as a memristor \cite{strukov2008missing,francesco2022}.  In particular, lowering the response threshold of a circuit by increasing interlayer coupling $g$ could be used to model long-term potentiation, which is a physiological mechanism that underlies learning.

{\it Acknowledgements}
We acknowledge valuable discussions with Onur Erten and Takashi Oka.
This work was supported by the Center for Emergent Materials, an NSF MRSEC, under
241 Grant No. DMR-201187
A. AlShafey and G. McCaul contributed equally to this work.


\bibliography{Ultrafast_laser-driven_dynamics_in_metal-insulator_interface}

\providecommand{\noopsort}[1]{}\providecommand{\singleletter}[1]{#1}%
\begin{thebibliography}{34}%
\makeatletter
\providecommand \@ifxundefined [1]{%
 \@ifx{#1\undefined}
}%
\providecommand \@ifnum [1]{%
 \ifnum #1\expandafter \@firstoftwo
 \else \expandafter \@secondoftwo
 \fi
}%
\providecommand \@ifx [1]{%
 \ifx #1\expandafter \@firstoftwo
 \else \expandafter \@secondoftwo
 \fi
}%
\providecommand \natexlab [1]{#1}%
\providecommand \enquote  [1]{``#1''}%
\providecommand \bibnamefont  [1]{#1}%
\providecommand \bibfnamefont [1]{#1}%
\providecommand \citenamefont [1]{#1}%
\providecommand \href@noop [0]{\@secondoftwo}%
\providecommand \href [0]{\begingroup \@sanitize@url \@href}%
\providecommand \@href[1]{\@@startlink{#1}\@@href}%
\providecommand \@@href[1]{\endgroup#1\@@endlink}%
\providecommand \@sanitize@url [0]{\catcode `\\12\catcode `\$12\catcode
  `\&12\catcode `\#12\catcode `\^12\catcode `\_12\catcode `\%12\relax}%
\providecommand \@@startlink[1]{}%
\providecommand \@@endlink[0]{}%
\providecommand \url  [0]{\begingroup\@sanitize@url \@url }%
\providecommand \@url [1]{\endgroup\@href {#1}{\urlprefix }}%
\providecommand \urlprefix  [0]{URL }%
\providecommand \Eprint [0]{\href }%
\providecommand \doibase [0]{https://doi.org/}%
\providecommand \selectlanguage [0]{\@gobble}%
\providecommand \bibinfo  [0]{\@secondoftwo}%
\providecommand \bibfield  [0]{\@secondoftwo}%
\providecommand \translation [1]{[#1]}%
\providecommand \BibitemOpen [0]{}%
\providecommand \bibitemStop [0]{}%
\providecommand \bibitemNoStop [0]{.\EOS\space}%
\providecommand \EOS [0]{\spacefactor3000\relax}%
\providecommand \BibitemShut  [1]{\csname bibitem#1\endcsname}%
\let\auto@bib@innerbib\@empty
\bibitem [{\citenamefont {Siegrist}\ \emph {et~al.}(2019)\citenamefont
  {Siegrist}, \citenamefont {Gessner}, \citenamefont {Ossiander}, \citenamefont
  {Denker}, \citenamefont {Chang}, \citenamefont {Schr{\"o}der}, \citenamefont
  {Guggenmos}, \citenamefont {Cui}, \citenamefont {Walowski}, \citenamefont
  {Martens} \emph {et~al.}}]{Siegrist:19}%
  \BibitemOpen
  \bibfield  {author} {\bibinfo {author} {\bibfnamefont {F.}~\bibnamefont
  {Siegrist}}, \bibinfo {author} {\bibfnamefont {J.~A.}\ \bibnamefont
  {Gessner}}, \bibinfo {author} {\bibfnamefont {M.}~\bibnamefont {Ossiander}},
  \bibinfo {author} {\bibfnamefont {C.}~\bibnamefont {Denker}}, \bibinfo
  {author} {\bibfnamefont {Y.-P.}\ \bibnamefont {Chang}}, \bibinfo {author}
  {\bibfnamefont {M.~C.}\ \bibnamefont {Schr{\"o}der}}, \bibinfo {author}
  {\bibfnamefont {A.}~\bibnamefont {Guggenmos}}, \bibinfo {author}
  {\bibfnamefont {Y.}~\bibnamefont {Cui}}, \bibinfo {author} {\bibfnamefont
  {J.}~\bibnamefont {Walowski}}, \bibinfo {author} {\bibfnamefont
  {U.}~\bibnamefont {Martens}}, \emph {et~al.},\ }\bibfield  {title} {\bibinfo
  {title} {Light-wave dynamic control of magnetism},\ }\href@noop {} {\bibfield
   {journal} {\bibinfo  {journal} {Nature}\ }\textbf {\bibinfo {volume}
  {571}},\ \bibinfo {pages} {240} (\bibinfo {year} {2019})}\BibitemShut
  {NoStop}%
\bibitem [{\citenamefont {Hellman}\ \emph {et~al.}(2017)\citenamefont
  {Hellman}, \citenamefont {Hoffmann}, \citenamefont {Tserkovnyak},
  \citenamefont {Beach}, \citenamefont {Fullerton}, \citenamefont {Leighton},
  \citenamefont {MacDonald}, \citenamefont {Ralph}, \citenamefont {Arena},
  \citenamefont {D{\"u}rr} \emph {et~al.}}]{hellman2017interface}%
  \BibitemOpen
  \bibfield  {author} {\bibinfo {author} {\bibfnamefont {F.}~\bibnamefont
  {Hellman}}, \bibinfo {author} {\bibfnamefont {A.}~\bibnamefont {Hoffmann}},
  \bibinfo {author} {\bibfnamefont {Y.}~\bibnamefont {Tserkovnyak}}, \bibinfo
  {author} {\bibfnamefont {G.~S.}\ \bibnamefont {Beach}}, \bibinfo {author}
  {\bibfnamefont {E.~E.}\ \bibnamefont {Fullerton}}, \bibinfo {author}
  {\bibfnamefont {C.}~\bibnamefont {Leighton}}, \bibinfo {author}
  {\bibfnamefont {A.~H.}\ \bibnamefont {MacDonald}}, \bibinfo {author}
  {\bibfnamefont {D.~C.}\ \bibnamefont {Ralph}}, \bibinfo {author}
  {\bibfnamefont {D.~A.}\ \bibnamefont {Arena}}, \bibinfo {author}
  {\bibfnamefont {H.~A.}\ \bibnamefont {D{\"u}rr}}, \emph {et~al.},\ }\bibfield
   {title} {\bibinfo {title} {Interface-induced phenomena in magnetism},\
  }\href@noop {} {\bibfield  {journal} {\bibinfo  {journal} {Reviews of modern
  physics}\ }\textbf {\bibinfo {volume} {89}},\ \bibinfo {pages} {025006}
  (\bibinfo {year} {2017})}\BibitemShut {NoStop}%
\bibitem [{\citenamefont {Kirilyuk}\ \emph {et~al.}(2010)\citenamefont
  {Kirilyuk}, \citenamefont {Kimel},\ and\ \citenamefont
  {Rasing}}]{kirilyuk2010ultrafast}%
  \BibitemOpen
  \bibfield  {author} {\bibinfo {author} {\bibfnamefont {A.}~\bibnamefont
  {Kirilyuk}}, \bibinfo {author} {\bibfnamefont {A.~V.}\ \bibnamefont
  {Kimel}},\ and\ \bibinfo {author} {\bibfnamefont {T.}~\bibnamefont
  {Rasing}},\ }\bibfield  {title} {\bibinfo {title} {Ultrafast optical
  manipulation of magnetic order},\ }\href@noop {} {\bibfield  {journal}
  {\bibinfo  {journal} {Reviews of Modern Physics}\ }\textbf {\bibinfo {volume}
  {82}},\ \bibinfo {pages} {2731} (\bibinfo {year} {2010})}\BibitemShut
  {NoStop}%
\bibitem [{\citenamefont {Strukov}\ \emph {et~al.}(2008)\citenamefont
  {Strukov}, \citenamefont {Snider}, \citenamefont {Stewart},\ and\
  \citenamefont {Williams}}]{strukov2008missing}%
  \BibitemOpen
  \bibfield  {author} {\bibinfo {author} {\bibfnamefont {D.~B.}\ \bibnamefont
  {Strukov}}, \bibinfo {author} {\bibfnamefont {G.~S.}\ \bibnamefont {Snider}},
  \bibinfo {author} {\bibfnamefont {D.~R.}\ \bibnamefont {Stewart}},\ and\
  \bibinfo {author} {\bibfnamefont {R.~S.}\ \bibnamefont {Williams}},\
  }\bibfield  {title} {\bibinfo {title} {The missing memristor found},\
  }\href@noop {} {\bibfield  {journal} {\bibinfo  {journal} {nature}\ }\textbf
  {\bibinfo {volume} {453}},\ \bibinfo {pages} {80} (\bibinfo {year}
  {2008})}\BibitemShut {NoStop}%
\bibitem [{\citenamefont {Heinrich}\ \emph {et~al.}(2011)\citenamefont
  {Heinrich}, \citenamefont {Burrowes}, \citenamefont {Montoya}, \citenamefont
  {Kardasz}, \citenamefont {Girt}, \citenamefont {Song}, \citenamefont {Sun},\
  and\ \citenamefont {Wu}}]{heinrich2011spin}%
  \BibitemOpen
  \bibfield  {author} {\bibinfo {author} {\bibfnamefont {B.}~\bibnamefont
  {Heinrich}}, \bibinfo {author} {\bibfnamefont {C.}~\bibnamefont {Burrowes}},
  \bibinfo {author} {\bibfnamefont {E.}~\bibnamefont {Montoya}}, \bibinfo
  {author} {\bibfnamefont {B.}~\bibnamefont {Kardasz}}, \bibinfo {author}
  {\bibfnamefont {E.}~\bibnamefont {Girt}}, \bibinfo {author} {\bibfnamefont
  {Y.-Y.}\ \bibnamefont {Song}}, \bibinfo {author} {\bibfnamefont
  {Y.}~\bibnamefont {Sun}},\ and\ \bibinfo {author} {\bibfnamefont
  {M.}~\bibnamefont {Wu}},\ }\bibfield  {title} {\bibinfo {title} {Spin pumping
  at the magnetic insulator (yig)/normal metal (au) interfaces},\ }\href@noop
  {} {\bibfield  {journal} {\bibinfo  {journal} {Physical review letters}\
  }\textbf {\bibinfo {volume} {107}},\ \bibinfo {pages} {066604} (\bibinfo
  {year} {2011})}\BibitemShut {NoStop}%
\bibitem [{\citenamefont {Ferray}\ \emph {et~al.}(1988)\citenamefont {Ferray},
  \citenamefont {L'Huillier}, \citenamefont {Li}, \citenamefont {Lompre},
  \citenamefont {Mainfray},\ and\ \citenamefont {Manus}}]{ferray1988multiple}%
  \BibitemOpen
  \bibfield  {author} {\bibinfo {author} {\bibfnamefont {M.}~\bibnamefont
  {Ferray}}, \bibinfo {author} {\bibfnamefont {A.}~\bibnamefont {L'Huillier}},
  \bibinfo {author} {\bibfnamefont {X.}~\bibnamefont {Li}}, \bibinfo {author}
  {\bibfnamefont {L.}~\bibnamefont {Lompre}}, \bibinfo {author} {\bibfnamefont
  {G.}~\bibnamefont {Mainfray}},\ and\ \bibinfo {author} {\bibfnamefont
  {C.}~\bibnamefont {Manus}},\ }\bibfield  {title} {\bibinfo {title}
  {Multiple-harmonic conversion of 1064 nm radiation in rare gases},\
  }\href@noop {} {\bibfield  {journal} {\bibinfo  {journal} {Journal of Physics
  B: Atomic, Molecular and Optical Physics}\ }\textbf {\bibinfo {volume}
  {21}},\ \bibinfo {pages} {L31} (\bibinfo {year} {1988})}\BibitemShut
  {NoStop}%
\bibitem [{\citenamefont {Ciappina}\ \emph {et~al.}(2015)\citenamefont
  {Ciappina}, \citenamefont {P\'erez-Hern\'andez}, \citenamefont {Landsman},
  \citenamefont {Zimmermann}, \citenamefont {Lewenstein}, \citenamefont
  {Roso},\ and\ \citenamefont {Krausz}}]{Ciappina2015PRL}%
  \BibitemOpen
  \bibfield  {author} {\bibinfo {author} {\bibfnamefont {M.~F.}\ \bibnamefont
  {Ciappina}}, \bibinfo {author} {\bibfnamefont {J.~A.}\ \bibnamefont
  {P\'erez-Hern\'andez}}, \bibinfo {author} {\bibfnamefont {A.~S.}\
  \bibnamefont {Landsman}}, \bibinfo {author} {\bibfnamefont {T.}~\bibnamefont
  {Zimmermann}}, \bibinfo {author} {\bibfnamefont {M.}~\bibnamefont
  {Lewenstein}}, \bibinfo {author} {\bibfnamefont {L.}~\bibnamefont {Roso}},\
  and\ \bibinfo {author} {\bibfnamefont {F.}~\bibnamefont {Krausz}},\
  }\bibfield  {title} {\bibinfo {title} {Carrier-wave rabi-flopping signatures
  in high-order harmonic generation for alkali atoms},\ }\href
  {https://doi.org/10.1103/PhysRevLett.114.143902} {\bibfield  {journal}
  {\bibinfo  {journal} {Phys. Rev. Lett.}\ }\textbf {\bibinfo {volume} {114}},\
  \bibinfo {pages} {143902} (\bibinfo {year} {2015})}\BibitemShut {NoStop}%
\bibitem [{\citenamefont {Li}\ \emph {et~al.}(2020)\citenamefont {Li},
  \citenamefont {Lu}, \citenamefont {Chew}, \citenamefont {Han}, \citenamefont
  {Li}, \citenamefont {Wu}, \citenamefont {Wang}, \citenamefont {Ghimire},\
  and\ \citenamefont {Chang}}]{li2020attosecond}%
  \BibitemOpen
  \bibfield  {author} {\bibinfo {author} {\bibfnamefont {J.}~\bibnamefont
  {Li}}, \bibinfo {author} {\bibfnamefont {J.}~\bibnamefont {Lu}}, \bibinfo
  {author} {\bibfnamefont {A.}~\bibnamefont {Chew}}, \bibinfo {author}
  {\bibfnamefont {S.}~\bibnamefont {Han}}, \bibinfo {author} {\bibfnamefont
  {J.}~\bibnamefont {Li}}, \bibinfo {author} {\bibfnamefont {Y.}~\bibnamefont
  {Wu}}, \bibinfo {author} {\bibfnamefont {H.}~\bibnamefont {Wang}}, \bibinfo
  {author} {\bibfnamefont {S.}~\bibnamefont {Ghimire}},\ and\ \bibinfo {author}
  {\bibfnamefont {Z.}~\bibnamefont {Chang}},\ }\bibfield  {title} {\bibinfo
  {title} {Attosecond science based on high harmonic generation from gases and
  solids},\ }\href@noop {} {\bibfield  {journal} {\bibinfo  {journal} {Nature
  Communications}\ }\textbf {\bibinfo {volume} {11}},\ \bibinfo {pages} {1}
  (\bibinfo {year} {2020})}\BibitemShut {NoStop}%
\bibitem [{\citenamefont {Ghimire}\ \emph {et~al.}(2011)\citenamefont
  {Ghimire}, \citenamefont {DiChiara}, \citenamefont {Sistrunk}, \citenamefont
  {Agostini}, \citenamefont {DiMauro},\ and\ \citenamefont
  {Reis}}]{ghimire2011observation}%
  \BibitemOpen
  \bibfield  {author} {\bibinfo {author} {\bibfnamefont {S.}~\bibnamefont
  {Ghimire}}, \bibinfo {author} {\bibfnamefont {A.~D.}\ \bibnamefont
  {DiChiara}}, \bibinfo {author} {\bibfnamefont {E.}~\bibnamefont {Sistrunk}},
  \bibinfo {author} {\bibfnamefont {P.}~\bibnamefont {Agostini}}, \bibinfo
  {author} {\bibfnamefont {L.~F.}\ \bibnamefont {DiMauro}},\ and\ \bibinfo
  {author} {\bibfnamefont {D.~A.}\ \bibnamefont {Reis}},\ }\bibfield  {title}
  {\bibinfo {title} {Observation of high-order harmonic generation in a bulk
  crystal},\ }\href@noop {} {\bibfield  {journal} {\bibinfo  {journal} {Nature
  physics}\ }\textbf {\bibinfo {volume} {7}},\ \bibinfo {pages} {138} (\bibinfo
  {year} {2011})}\BibitemShut {NoStop}%
\bibitem [{\citenamefont {Hohenleutner}\ \emph
  {et~al.}(2015{\natexlab{a}})\citenamefont {Hohenleutner}, \citenamefont
  {Langer}, \citenamefont {Schubert}, \citenamefont {Knorr}, \citenamefont
  {Huttner}, \citenamefont {Koch}, \citenamefont {Kira},\ and\ \citenamefont
  {Huber}}]{hohenleutner2015real}%
  \BibitemOpen
  \bibfield  {author} {\bibinfo {author} {\bibfnamefont {M.}~\bibnamefont
  {Hohenleutner}}, \bibinfo {author} {\bibfnamefont {F.}~\bibnamefont
  {Langer}}, \bibinfo {author} {\bibfnamefont {O.}~\bibnamefont {Schubert}},
  \bibinfo {author} {\bibfnamefont {M.}~\bibnamefont {Knorr}}, \bibinfo
  {author} {\bibfnamefont {U.}~\bibnamefont {Huttner}}, \bibinfo {author}
  {\bibfnamefont {S.~W.}\ \bibnamefont {Koch}}, \bibinfo {author}
  {\bibfnamefont {M.}~\bibnamefont {Kira}},\ and\ \bibinfo {author}
  {\bibfnamefont {R.}~\bibnamefont {Huber}},\ }\bibfield  {title} {\bibinfo
  {title} {Real-time observation of interfering crystal electrons in
  high-harmonic generation},\ }\href@noop {} {\bibfield  {journal} {\bibinfo
  {journal} {Nature}\ }\textbf {\bibinfo {volume} {523}},\ \bibinfo {pages}
  {572} (\bibinfo {year} {2015}{\natexlab{a}})}\BibitemShut {NoStop}%
\bibitem [{\citenamefont {Luu}\ \emph {et~al.}(2015)\citenamefont {Luu},
  \citenamefont {Garg}, \citenamefont {Kruchinin}, \citenamefont {Moulet},
  \citenamefont {Hassan},\ and\ \citenamefont {Goulielmakis}}]{luu2015extreme}%
  \BibitemOpen
  \bibfield  {author} {\bibinfo {author} {\bibfnamefont {T.~T.}\ \bibnamefont
  {Luu}}, \bibinfo {author} {\bibfnamefont {M.}~\bibnamefont {Garg}}, \bibinfo
  {author} {\bibfnamefont {S.~Y.}\ \bibnamefont {Kruchinin}}, \bibinfo {author}
  {\bibfnamefont {A.}~\bibnamefont {Moulet}}, \bibinfo {author} {\bibfnamefont
  {M.~T.}\ \bibnamefont {Hassan}},\ and\ \bibinfo {author} {\bibfnamefont
  {E.}~\bibnamefont {Goulielmakis}},\ }\bibfield  {title} {\bibinfo {title}
  {Extreme ultraviolet high-harmonic spectroscopy of solids},\ }\href@noop {}
  {\bibfield  {journal} {\bibinfo  {journal} {Nature}\ }\textbf {\bibinfo
  {volume} {521}},\ \bibinfo {pages} {498} (\bibinfo {year}
  {2015})}\BibitemShut {NoStop}%
\bibitem [{\citenamefont {Ortmann}\ \emph {et~al.}(2017)\citenamefont
  {Ortmann}, \citenamefont {P\'erez-Hern\'andez}, \citenamefont {Ciappina},
  \citenamefont {Sch\"otz}, \citenamefont {Chac\'on}, \citenamefont {Zeraouli},
  \citenamefont {Kling}, \citenamefont {Roso}, \citenamefont {Lewenstein},\
  and\ \citenamefont {Landsman}}]{Lisa2017PRL}%
  \BibitemOpen
  \bibfield  {author} {\bibinfo {author} {\bibfnamefont {L.}~\bibnamefont
  {Ortmann}}, \bibinfo {author} {\bibfnamefont {J.~A.}\ \bibnamefont
  {P\'erez-Hern\'andez}}, \bibinfo {author} {\bibfnamefont {M.~F.}\
  \bibnamefont {Ciappina}}, \bibinfo {author} {\bibfnamefont {J.}~\bibnamefont
  {Sch\"otz}}, \bibinfo {author} {\bibfnamefont {A.}~\bibnamefont {Chac\'on}},
  \bibinfo {author} {\bibfnamefont {G.}~\bibnamefont {Zeraouli}}, \bibinfo
  {author} {\bibfnamefont {M.~F.}\ \bibnamefont {Kling}}, \bibinfo {author}
  {\bibfnamefont {L.}~\bibnamefont {Roso}}, \bibinfo {author} {\bibfnamefont
  {M.}~\bibnamefont {Lewenstein}},\ and\ \bibinfo {author} {\bibfnamefont
  {A.~S.}\ \bibnamefont {Landsman}},\ }\bibfield  {title} {\bibinfo {title}
  {Emergence of a higher energy structure in strong field ionization with
  inhomogeneous electric fields},\ }\href
  {https://doi.org/10.1103/PhysRevLett.119.053204} {\bibfield  {journal}
  {\bibinfo  {journal} {Phys. Rev. Lett.}\ }\textbf {\bibinfo {volume} {119}},\
  \bibinfo {pages} {053204} (\bibinfo {year} {2017})}\BibitemShut {NoStop}%
\bibitem [{\citenamefont {Vampa}\ \emph {et~al.}(2017)\citenamefont {Vampa},
  \citenamefont {Ghamsari}, \citenamefont {Siadat~Mousavi}, \citenamefont
  {Hammond}, \citenamefont {Olivieri}, \citenamefont {Lisicka-Skrek},
  \citenamefont {Naumov}, \citenamefont {Villeneuve}, \citenamefont {Staudte},
  \citenamefont {Berini} \emph {et~al.}}]{Vampa2017}%
  \BibitemOpen
  \bibfield  {author} {\bibinfo {author} {\bibfnamefont {G.}~\bibnamefont
  {Vampa}}, \bibinfo {author} {\bibfnamefont {B.}~\bibnamefont {Ghamsari}},
  \bibinfo {author} {\bibfnamefont {S.}~\bibnamefont {Siadat~Mousavi}},
  \bibinfo {author} {\bibfnamefont {T.}~\bibnamefont {Hammond}}, \bibinfo
  {author} {\bibfnamefont {A.}~\bibnamefont {Olivieri}}, \bibinfo {author}
  {\bibfnamefont {E.}~\bibnamefont {Lisicka-Skrek}}, \bibinfo {author}
  {\bibfnamefont {A.~Y.}\ \bibnamefont {Naumov}}, \bibinfo {author}
  {\bibfnamefont {D.}~\bibnamefont {Villeneuve}}, \bibinfo {author}
  {\bibfnamefont {A.}~\bibnamefont {Staudte}}, \bibinfo {author} {\bibfnamefont
  {P.}~\bibnamefont {Berini}}, \emph {et~al.},\ }\bibfield  {title} {\bibinfo
  {title} {Plasmon-enhanced high-harmonic generation from silicon},\
  }\href@noop {} {\bibfield  {journal} {\bibinfo  {journal} {Nature Physics}\
  }\textbf {\bibinfo {volume} {13}},\ \bibinfo {pages} {659} (\bibinfo {year}
  {2017})}\BibitemShut {NoStop}%
\bibitem [{\citenamefont {Vampa}\ \emph {et~al.}(2014)\citenamefont {Vampa},
  \citenamefont {McDonald}, \citenamefont {Orlando}, \citenamefont {Klug},
  \citenamefont {Corkum},\ and\ \citenamefont {Brabec}}]{vampa2014theoretical}%
  \BibitemOpen
  \bibfield  {author} {\bibinfo {author} {\bibfnamefont {G.}~\bibnamefont
  {Vampa}}, \bibinfo {author} {\bibfnamefont {C.}~\bibnamefont {McDonald}},
  \bibinfo {author} {\bibfnamefont {G.}~\bibnamefont {Orlando}}, \bibinfo
  {author} {\bibfnamefont {D.}~\bibnamefont {Klug}}, \bibinfo {author}
  {\bibfnamefont {P.}~\bibnamefont {Corkum}},\ and\ \bibinfo {author}
  {\bibfnamefont {T.}~\bibnamefont {Brabec}},\ }\bibfield  {title} {\bibinfo
  {title} {Theoretical analysis of high-harmonic generation in solids},\
  }\href@noop {} {\bibfield  {journal} {\bibinfo  {journal} {Physical review
  letters}\ }\textbf {\bibinfo {volume} {113}},\ \bibinfo {pages} {073901}
  (\bibinfo {year} {2014})}\BibitemShut {NoStop}%
\bibitem [{\citenamefont {Ghimire}\ and\ \citenamefont
  {Reis}(2019)}]{ghimire2019high}%
  \BibitemOpen
  \bibfield  {author} {\bibinfo {author} {\bibfnamefont {S.}~\bibnamefont
  {Ghimire}}\ and\ \bibinfo {author} {\bibfnamefont {D.~A.}\ \bibnamefont
  {Reis}},\ }\bibfield  {title} {\bibinfo {title} {High-harmonic generation
  from solids},\ }\href@noop {} {\bibfield  {journal} {\bibinfo  {journal}
  {Nature physics}\ }\textbf {\bibinfo {volume} {15}},\ \bibinfo {pages} {10}
  (\bibinfo {year} {2019})}\BibitemShut {NoStop}%
\bibitem [{\citenamefont {Silva}\ \emph {et~al.}(2018)\citenamefont {Silva},
  \citenamefont {Blinov}, \citenamefont {Rubtsov}, \citenamefont {Smirnova},\
  and\ \citenamefont {Ivanov}}]{Silva:18}%
  \BibitemOpen
  \bibfield  {author} {\bibinfo {author} {\bibfnamefont {R.}~\bibnamefont
  {Silva}}, \bibinfo {author} {\bibfnamefont {I.~V.}\ \bibnamefont {Blinov}},
  \bibinfo {author} {\bibfnamefont {A.~N.}\ \bibnamefont {Rubtsov}}, \bibinfo
  {author} {\bibfnamefont {O.}~\bibnamefont {Smirnova}},\ and\ \bibinfo
  {author} {\bibfnamefont {M.}~\bibnamefont {Ivanov}},\ }\bibfield  {title}
  {\bibinfo {title} {High-harmonic spectroscopy of ultrafast many-body dynamics
  in strongly correlated systems},\ }\href@noop {} {\bibfield  {journal}
  {\bibinfo  {journal} {Nature Photonics}\ }\textbf {\bibinfo {volume} {12}},\
  \bibinfo {pages} {266} (\bibinfo {year} {2018})}\BibitemShut {NoStop}%
\bibitem [{\citenamefont {Murakami}\ \emph {et~al.}(2021)\citenamefont
  {Murakami}, \citenamefont {Takayoshi}, \citenamefont {Koga},\ and\
  \citenamefont {Werner}}]{murakami2021high}%
  \BibitemOpen
  \bibfield  {author} {\bibinfo {author} {\bibfnamefont {Y.}~\bibnamefont
  {Murakami}}, \bibinfo {author} {\bibfnamefont {S.}~\bibnamefont {Takayoshi}},
  \bibinfo {author} {\bibfnamefont {A.}~\bibnamefont {Koga}},\ and\ \bibinfo
  {author} {\bibfnamefont {P.}~\bibnamefont {Werner}},\ }\bibfield  {title}
  {\bibinfo {title} {High-harmonic generation in one-dimensional mott
  insulators},\ }\href@noop {} {\bibfield  {journal} {\bibinfo  {journal}
  {Physical Review B}\ }\textbf {\bibinfo {volume} {103}},\ \bibinfo {pages}
  {035110} (\bibinfo {year} {2021})}\BibitemShut {NoStop}%
\bibitem [{\citenamefont {Orthodoxou}\ \emph {et~al.}(2021)\citenamefont
  {Orthodoxou}, \citenamefont {Za{\"\i}r},\ and\ \citenamefont
  {Booth}}]{orthodoxou2021high}%
  \BibitemOpen
  \bibfield  {author} {\bibinfo {author} {\bibfnamefont {C.}~\bibnamefont
  {Orthodoxou}}, \bibinfo {author} {\bibfnamefont {A.}~\bibnamefont
  {Za{\"\i}r}},\ and\ \bibinfo {author} {\bibfnamefont {G.~H.}\ \bibnamefont
  {Booth}},\ }\bibfield  {title} {\bibinfo {title} {High harmonic generation in
  two-dimensional mott insulators},\ }\href@noop {} {\bibfield  {journal}
  {\bibinfo  {journal} {arXiv preprint arXiv:2103.13708}\ } (\bibinfo {year}
  {2021})}\BibitemShut {NoStop}%
\bibitem [{\citenamefont {McCaul}\ \emph
  {et~al.}(2020{\natexlab{a}})\citenamefont {McCaul}, \citenamefont
  {Orthodoxou}, \citenamefont {Jacobs}, \citenamefont {Booth},\ and\
  \citenamefont {Bondar}}]{tracking1}%
  \BibitemOpen
  \bibfield  {author} {\bibinfo {author} {\bibfnamefont {G.}~\bibnamefont
  {McCaul}}, \bibinfo {author} {\bibfnamefont {C.}~\bibnamefont {Orthodoxou}},
  \bibinfo {author} {\bibfnamefont {K.}~\bibnamefont {Jacobs}}, \bibinfo
  {author} {\bibfnamefont {G.~H.}\ \bibnamefont {Booth}},\ and\ \bibinfo
  {author} {\bibfnamefont {D.~I.}\ \bibnamefont {Bondar}},\ }\bibfield  {title}
  {\bibinfo {title} {Driven imposters: Controlling expectations in many-body
  systems},\ }\href {https://doi.org/10.1103/PhysRevLett.124.183201} {\bibfield
   {journal} {\bibinfo  {journal} {Phys. Rev. Lett.}\ }\textbf {\bibinfo
  {volume} {124}},\ \bibinfo {pages} {183201} (\bibinfo {year}
  {2020}{\natexlab{a}})}\BibitemShut {NoStop}%
\bibitem [{\citenamefont {McCaul}\ \emph
  {et~al.}(2020{\natexlab{b}})\citenamefont {McCaul}, \citenamefont
  {Orthodoxou}, \citenamefont {Jacobs}, \citenamefont {Booth},\ and\
  \citenamefont {Bondar}}]{tracking2}%
  \BibitemOpen
  \bibfield  {author} {\bibinfo {author} {\bibfnamefont {G.}~\bibnamefont
  {McCaul}}, \bibinfo {author} {\bibfnamefont {C.}~\bibnamefont {Orthodoxou}},
  \bibinfo {author} {\bibfnamefont {K.}~\bibnamefont {Jacobs}}, \bibinfo
  {author} {\bibfnamefont {G.~H.}\ \bibnamefont {Booth}},\ and\ \bibinfo
  {author} {\bibfnamefont {D.~I.}\ \bibnamefont {Bondar}},\ }\bibfield  {title}
  {\bibinfo {title} {Controlling arbitrary observables in correlated many-body
  systems},\ }\href {https://doi.org/10.1103/PhysRevA.101.053408} {\bibfield
  {journal} {\bibinfo  {journal} {Phys. Rev. A}\ }\textbf {\bibinfo {volume}
  {101}},\ \bibinfo {pages} {053408} (\bibinfo {year}
  {2020}{\natexlab{b}})}\BibitemShut {NoStop}%
\bibitem [{\citenamefont {McCaul}\ and\ \citenamefont
  {Bondar}(2021)}]{2104.06322}%
  \BibitemOpen
  \bibfield  {author} {\bibinfo {author} {\bibfnamefont {G.}~\bibnamefont
  {McCaul}}\ and\ \bibinfo {author} {\bibfnamefont {D.~I.}\ \bibnamefont
  {Bondar}},\ }\href@noop {} {\bibinfo {title} {Towards single atom computing
  via high harmonic generation}} (\bibinfo {year} {2021}),\ \Eprint
  {https://arxiv.org/abs/arXiv:2104.06322} {arXiv:2104.06322} \BibitemShut
  {NoStop}%
\bibitem [{\citenamefont {Silva}\ \emph {et~al.}(2019)\citenamefont {Silva},
  \citenamefont {Jim{\'e}nez-Gal{\'a}n}, \citenamefont {Amorim}, \citenamefont
  {Smirnova},\ and\ \citenamefont {Ivanov}}]{silva2019topological}%
  \BibitemOpen
  \bibfield  {author} {\bibinfo {author} {\bibfnamefont {R.}~\bibnamefont
  {Silva}}, \bibinfo {author} {\bibfnamefont {{\'A}.}~\bibnamefont
  {Jim{\'e}nez-Gal{\'a}n}}, \bibinfo {author} {\bibfnamefont {B.}~\bibnamefont
  {Amorim}}, \bibinfo {author} {\bibfnamefont {O.}~\bibnamefont {Smirnova}},\
  and\ \bibinfo {author} {\bibfnamefont {M.}~\bibnamefont {Ivanov}},\
  }\bibfield  {title} {\bibinfo {title} {Topological strong-field physics on
  sub-laser-cycle timescale},\ }\href@noop {} {\bibfield  {journal} {\bibinfo
  {journal} {Nature Photonics}\ }\textbf {\bibinfo {volume} {13}},\ \bibinfo
  {pages} {849} (\bibinfo {year} {2019})}\BibitemShut {NoStop}%
\bibitem [{\citenamefont {Magann}\ \emph {et~al.}(2022)\citenamefont {Magann},
  \citenamefont {McCaul}, \citenamefont {Rabitz},\ and\ \citenamefont
  {Bondar}}]{mixing}%
  \BibitemOpen
  \bibfield  {author} {\bibinfo {author} {\bibfnamefont {A.~B.}\ \bibnamefont
  {Magann}}, \bibinfo {author} {\bibfnamefont {G.}~\bibnamefont {McCaul}},
  \bibinfo {author} {\bibfnamefont {H.~A.}\ \bibnamefont {Rabitz}},\ and\
  \bibinfo {author} {\bibfnamefont {D.~I.}\ \bibnamefont {Bondar}},\ }\bibfield
   {title} {\bibinfo {title} {Sequential optical response suppression for
  chemical mixture characterization},\ }\href@noop {} {\bibfield  {journal}
  {\bibinfo  {journal} {Quantum}\ }\textbf {\bibinfo {volume} {6}},\ \bibinfo
  {pages} {626} (\bibinfo {year} {2022})}\BibitemShut {NoStop}%
\bibitem [{\citenamefont {Oka}(2012{\natexlab{a}})}]{PhysRevB.86.075148}%
  \BibitemOpen
  \bibfield  {author} {\bibinfo {author} {\bibfnamefont {T.}~\bibnamefont
  {Oka}},\ }\bibfield  {title} {\bibinfo {title} {Nonlinear doublon production
  in a mott insulator: Landau-dykhne method applied to an integrable model},\
  }\href {https://doi.org/10.1103/PhysRevB.86.075148} {\bibfield  {journal}
  {\bibinfo  {journal} {Phys. Rev. B}\ }\textbf {\bibinfo {volume} {86}},\
  \bibinfo {pages} {075148} (\bibinfo {year} {2012}{\natexlab{a}})}\BibitemShut
  {NoStop}%
\bibitem [{\citenamefont {McCaul}\ \emph
  {et~al.}(2020{\natexlab{c}})\citenamefont {McCaul}, \citenamefont
  {Orthodoxou}, \citenamefont {Jacobs}, \citenamefont {Booth},\ and\
  \citenamefont {Bondar}}]{mccaul2020driven}%
  \BibitemOpen
  \bibfield  {author} {\bibinfo {author} {\bibfnamefont {G.}~\bibnamefont
  {McCaul}}, \bibinfo {author} {\bibfnamefont {C.}~\bibnamefont {Orthodoxou}},
  \bibinfo {author} {\bibfnamefont {K.}~\bibnamefont {Jacobs}}, \bibinfo
  {author} {\bibfnamefont {G.~H.}\ \bibnamefont {Booth}},\ and\ \bibinfo
  {author} {\bibfnamefont {D.~I.}\ \bibnamefont {Bondar}},\ }\bibfield  {title}
  {\bibinfo {title} {Driven imposters: Controlling expectations in many-body
  systems},\ }\href@noop {} {\bibfield  {journal} {\bibinfo  {journal}
  {Physical Review Letters}\ }\textbf {\bibinfo {volume} {124}},\ \bibinfo
  {pages} {183201} (\bibinfo {year} {2020}{\natexlab{c}})}\BibitemShut
  {NoStop}%
\bibitem [{\citenamefont {Leigh}\ and\ \citenamefont
  {Phillips}(2009)}]{PhysRevB.79.245120}%
  \BibitemOpen
  \bibfield  {author} {\bibinfo {author} {\bibfnamefont {R.~G.}\ \bibnamefont
  {Leigh}}\ and\ \bibinfo {author} {\bibfnamefont {P.}~\bibnamefont
  {Phillips}},\ }\bibfield  {title} {\bibinfo {title} {Origin of the mott
  gap},\ }\href {https://doi.org/10.1103/PhysRevB.79.245120} {\bibfield
  {journal} {\bibinfo  {journal} {Phys. Rev. B}\ }\textbf {\bibinfo {volume}
  {79}},\ \bibinfo {pages} {245120} (\bibinfo {year} {2009})}\BibitemShut
  {NoStop}%
\bibitem [{\citenamefont {Sunko}\ \emph {et~al.}(2020)\citenamefont {Sunko},
  \citenamefont {Mazzola}, \citenamefont {Kitamura}, \citenamefont {Khim},
  \citenamefont {Kushwaha}, \citenamefont {Clark}, \citenamefont {Watson},
  \citenamefont {Markovi{\'c}}, \citenamefont {Biswas}, \citenamefont
  {Pourovskii} \emph {et~al.}}]{Takashi:20}%
  \BibitemOpen
  \bibfield  {author} {\bibinfo {author} {\bibfnamefont {V.}~\bibnamefont
  {Sunko}}, \bibinfo {author} {\bibfnamefont {F.}~\bibnamefont {Mazzola}},
  \bibinfo {author} {\bibfnamefont {S.}~\bibnamefont {Kitamura}}, \bibinfo
  {author} {\bibfnamefont {S.}~\bibnamefont {Khim}}, \bibinfo {author}
  {\bibfnamefont {P.}~\bibnamefont {Kushwaha}}, \bibinfo {author}
  {\bibfnamefont {O.}~\bibnamefont {Clark}}, \bibinfo {author} {\bibfnamefont
  {M.}~\bibnamefont {Watson}}, \bibinfo {author} {\bibfnamefont
  {I.}~\bibnamefont {Markovi{\'c}}}, \bibinfo {author} {\bibfnamefont
  {D.}~\bibnamefont {Biswas}}, \bibinfo {author} {\bibfnamefont
  {L.}~\bibnamefont {Pourovskii}}, \emph {et~al.},\ }\bibfield  {title}
  {\bibinfo {title} {Probing spin correlations using angle-resolved
  photoemission in a coupled metallic/mott insulator system},\ }\href@noop {}
  {\bibfield  {journal} {\bibinfo  {journal} {Science advances}\ }\textbf
  {\bibinfo {volume} {6}},\ \bibinfo {pages} {eaaz0611} (\bibinfo {year}
  {2020})}\BibitemShut {NoStop}%
\bibitem [{\citenamefont {Abdelwahab}\ \emph {et~al.}(2015)\citenamefont
  {Abdelwahab}, \citenamefont {Jeckelmann},\ and\ \citenamefont
  {Hohenadler}}]{abdelwahab2015ground}%
  \BibitemOpen
  \bibfield  {author} {\bibinfo {author} {\bibfnamefont {A.}~\bibnamefont
  {Abdelwahab}}, \bibinfo {author} {\bibfnamefont {E.}~\bibnamefont
  {Jeckelmann}},\ and\ \bibinfo {author} {\bibfnamefont {M.}~\bibnamefont
  {Hohenadler}},\ }\bibfield  {title} {\bibinfo {title} {Ground-state and
  spectral properties of an asymmetric hubbard ladder},\ }\href@noop {}
  {\bibfield  {journal} {\bibinfo  {journal} {Physical Review B}\ }\textbf
  {\bibinfo {volume} {91}},\ \bibinfo {pages} {155119} (\bibinfo {year}
  {2015})}\BibitemShut {NoStop}%
\bibitem [{\citenamefont {Abdelwahab}\ and\ \citenamefont
  {Jeckelmann}(2018)}]{abdelwahab2018correlations}%
  \BibitemOpen
  \bibfield  {author} {\bibinfo {author} {\bibfnamefont {A.}~\bibnamefont
  {Abdelwahab}}\ and\ \bibinfo {author} {\bibfnamefont {E.}~\bibnamefont
  {Jeckelmann}},\ }\bibfield  {title} {\bibinfo {title} {Correlations and
  confinement of excitations in an asymmetric hubbard ladder},\ }\href@noop {}
  {\bibfield  {journal} {\bibinfo  {journal} {The European Physical Journal B}\
  }\textbf {\bibinfo {volume} {91}},\ \bibinfo {pages} {1} (\bibinfo {year}
  {2018})}\BibitemShut {NoStop}%
\bibitem [{\citenamefont {Hohenleutner}\ \emph
  {et~al.}(2015{\natexlab{b}})\citenamefont {Hohenleutner}, \citenamefont
  {Langer}, \citenamefont {Schubert}, \citenamefont {Knorr}, \citenamefont
  {Huttner}, \citenamefont {Koch}, \citenamefont {Kira},\ and\ \citenamefont
  {Huber}}]{Hohenleutner2015}%
  \BibitemOpen
  \bibfield  {author} {\bibinfo {author} {\bibfnamefont {M.}~\bibnamefont
  {Hohenleutner}}, \bibinfo {author} {\bibfnamefont {F.}~\bibnamefont
  {Langer}}, \bibinfo {author} {\bibfnamefont {O.}~\bibnamefont {Schubert}},
  \bibinfo {author} {\bibfnamefont {M.}~\bibnamefont {Knorr}}, \bibinfo
  {author} {\bibfnamefont {U.}~\bibnamefont {Huttner}}, \bibinfo {author}
  {\bibfnamefont {S.~W.}\ \bibnamefont {Koch}}, \bibinfo {author}
  {\bibfnamefont {M.}~\bibnamefont {Kira}},\ and\ \bibinfo {author}
  {\bibfnamefont {R.}~\bibnamefont {Huber}},\ }\bibfield  {title} {\bibinfo
  {title} {Real-time observation of interfering crystal electrons in
  high-harmonic generation},\ }\href@noop {} {\bibfield  {journal} {\bibinfo
  {journal} {Nature}\ }\textbf {\bibinfo {volume} {523}},\ \bibinfo {pages}
  {572 EP } (\bibinfo {year} {2015}{\natexlab{b}})}\BibitemShut {NoStop}%
\bibitem [{\citenamefont {Gebhard}(2010)}]{floriangebhard2010}%
  \BibitemOpen
  \bibfield  {author} {\bibinfo {author} {\bibfnamefont {F.}~\bibnamefont
  {Gebhard}},\ }\href {https://www.xarg.org/ref/a/3642082637/} {\emph {\bibinfo
  {title} {The Mott Metal-Insulator Transition: Models and Methods (Springer
  Tracts in Modern Physics)}}}\ (\bibinfo  {publisher} {Springer},\ \bibinfo
  {year} {2010})\BibitemShut {NoStop}%
\bibitem [{\citenamefont {Weinberg}\ and\ \citenamefont
  {Bukov}(2019)}]{quspin}%
  \BibitemOpen
  \bibfield  {author} {\bibinfo {author} {\bibfnamefont {P.}~\bibnamefont
  {Weinberg}}\ and\ \bibinfo {author} {\bibfnamefont {M.}~\bibnamefont
  {Bukov}},\ }\bibfield  {title} {\bibinfo {title} {{QuSpin: a Python Package
  for Dynamics and Exact Diagonalisation of Quantum Many Body Systems. Part II:
  bosons, fermions and higher spins}},\ }\href
  {https://doi.org/10.21468/SciPostPhys.7.2.020} {\bibfield  {journal}
  {\bibinfo  {journal} {SciPost Phys.}\ }\textbf {\bibinfo {volume} {7}},\
  \bibinfo {pages} {20} (\bibinfo {year} {2019})}\BibitemShut {NoStop}%
\bibitem [{\citenamefont {Oka}(2012{\natexlab{b}})}]{oka2012nonlinear}%
  \BibitemOpen
  \bibfield  {author} {\bibinfo {author} {\bibfnamefont {T.}~\bibnamefont
  {Oka}},\ }\bibfield  {title} {\bibinfo {title} {Nonlinear doublon production
  in a mott insulator: Landau-dykhne method applied to an integrable model},\
  }\href@noop {} {\bibfield  {journal} {\bibinfo  {journal} {Physical Review
  B}\ }\textbf {\bibinfo {volume} {86}},\ \bibinfo {pages} {075148} (\bibinfo
  {year} {2012}{\natexlab{b}})}\BibitemShut {NoStop}%
\bibitem [{\citenamefont {Peronaci}\ \emph {et~al.}(2021)\citenamefont
  {Peronaci}, \citenamefont {Ameli}, \citenamefont {Takayoshi}, \citenamefont
  {Landsman},\ and\ \citenamefont {Oka}}]{francesco2022}%
  \BibitemOpen
  \bibfield  {author} {\bibinfo {author} {\bibfnamefont {F.}~\bibnamefont
  {Peronaci}}, \bibinfo {author} {\bibfnamefont {S.}~\bibnamefont {Ameli}},
  \bibinfo {author} {\bibfnamefont {S.}~\bibnamefont {Takayoshi}}, \bibinfo
  {author} {\bibfnamefont {A.}~\bibnamefont {Landsman}},\ and\ \bibinfo
  {author} {\bibfnamefont {T.}~\bibnamefont {Oka}},\ }\bibfield  {title}
  {\bibinfo {title} {Mott memristors based on field-induced carrier avalanche
  multiplication},\ }\href@noop {} {\bibfield  {journal} {\bibinfo  {journal}
  {arXiv preprint arXiv:2104.00559}\ } (\bibinfo {year} {2021})}\BibitemShut
  {NoStop}%
\end{thebibliography}%
\end{document}